# Optical Vortex Transfer and Dispersion-Controlled Light Propagation in an Er$^{3+}$: YAG Three-Level Quantum System

**Arefeh Vaezi[1], Ali Mortezapour[1], Seyed Hossein Asadpour[2]**

[1]Department of Physics, University of Guilan, P.O. Box 41335–1914, Rasht, Iran.
[2]School of Quantum Physics and Matter, Institute for Research in Fundamental Sciences, IPM, Tehran, 19538-33511, Iran
e-mail: mortezapour@guilan.ac.ir (corresponding author)

**Abstract**

We theoretically investigate coherent orbital-angular-momentum (OAM) transfer and dispersion-controlled light propagation in a ladder-type Er$^{3+}$: YAG three-level system. Using the density-matrix formalism and coupled Maxwell–Bloch equations, we derive analytical expressions for the probe and generated beams that explicitly incorporate Er$^{3+}$-ion concentration. We show that an incident vortex-carrying probe beam transfers its OAM to a generated signal beam through a concentration-dependent sum-frequency nonlinear process, with complete phase and topological-charge preservation. By analyzing conversion efficiency and spatial phase and intensity distributions, we identify an optimal Er$^{3+}$ concentration (3%) that maximizes vortex-transfer efficiency. Furthermore, the absorption and dispersion spectra of the probe and generated beams reveal the mechanism underpinning the vortex transfer and demonstrate tunable transitions between fast and slow regimes. These results establish Er$^{3+}$:YAG as a viable solid-state platform for coherent manipulation of structured light, enabling efficient vortex-beam frequency conversion and dispersion engineering for applications in high-dimensional quantum communication, wavelength-compatible OAM interfaces, and slow-light photonic signal processing.

## 1. Introduction

Coherent light–matter interaction has become central to quantum optics due to applications in quantum information processing [1,2] and precision metrology [3,4]. Recently, increasing attention has been devoted to the interaction of structured light with atomic and solid-state systems. An essential class of structured light is the optical vortex beam, characterized by a helical phase factor $e^{i\ell\phi}$, where $\ell$ is the topological charge [5]. Such beams possess a phase singularity at the center, producing a doughnut-shaped intensity profile [5].

Optical vortex beams carry orbital angular momentum (OAM) along their propagation direction and exhibit unique interactions with matter. The concept was theoretically established in 1989 when Coullet et al. identified vortex solutions to the Maxwell–Bloch equations [6], and later experimentally verified when Allen et al. showed that laser beams can carry OAM [7]. Since then, OAM beams have been applied in quantum optics, information encoding [8,9], and optical manipulation [10], enabling phenomena such as atomic vortex beams [11, 12], spatially dependent electromagnetically induced transparency (EIT) [13,14], light-induced torque [15], and OAM-entangled photon pairs [16]. The ability to convert a vortex beam into a non-vortex beam, or to transfer OAM between fields, is highly relevant for topological photonics and quantum communication [17]. Such a mechanism is crucial as it enables the production of structured light at wavelengths where direct generation using standard optics is impossible (e.g., far infrared or ultraviolet). Moreover, Vortex exchange enables the generation of structured light at wavelengths inaccessible by direct optical methods and may assist in creating synthetic gauge fields for ultracold atoms [18].

On the other hand, the properties of the matter with which the light interacts can play a significant role in achieving the desired result or facilitating a new phenomenon. Solid-state platforms that simultaneously support long-lived coherence, large optical nonlinearities, and controllable dispersion remain of particular interest. Rare-earth-doped crystals offer these capabilities through narrow homogeneous linewidths, long optical and spin coherence times, and compatibility with low-power coherent excitation [19–21]. Among these systems, $Er^{3+}$-doped yttrium aluminum garnet ($Er^{3+}$: YAG) crystals have emerged as highly versatile media for exploring light–matter coherence effects. Their long-lived excited states, wide transparency window [22], and tunable optical properties [23] make them suitable for investigating electromagnetically induced transparency (EIT) [24], optical bistability [25–27], induced gratings [28, 29], and subluminal/superluminal propagation [30]. Notably, the $Er^{3+}$ concentration significantly plays a decisive role in determining the dipole moment, spontaneous decay rate, and Rabi frequencies, thereby modulating the nonlinear optical response [31]. Consequently, controlling the doping concentration provides a powerful means to manipulate dispersion and absorption, thereby engineering light propagation characteristics within the crystal.

In this work, we theoretically investigate vortex transfer and dispersion control in a three-level ladder-type $Er^{3+}$: YAG system. We show that a vortex-carrying probe beam transfers its OAM to a generated signal through a sum-frequency process, and we analyze how $Er^{3+}$ concentration governs conversion efficiency and dispersive light propagation.

The paper is organized as follows. In Sec. 2, we introduce the model and present the formulation of the basic set of equations by analytically solving the coupled Maxwell-Bloch equations. The results are presented in Section 3, while Section 4 summarizes the main findings.

## 2. Theoretical Model and Formulation

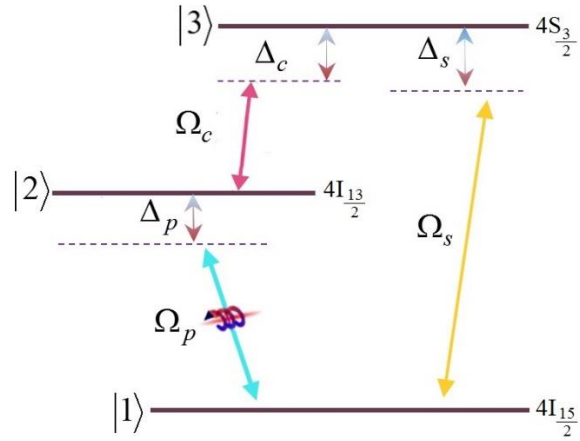

**Fig. 1.** Schematic depiction of the interaction of a three-level $Er^{3+}$ ion quantum system, which can carry optical vortices. In the proposed scheme, a weak probe field $E_p$ couples the $|1\rangle \leftrightarrow |2\rangle$ transition, and a strong control field $E_c$ couples the $|2\rangle \leftrightarrow |3\rangle$ transition, thereby a sum-frequency signal $E_s$ generated via a second-order nonlinear process.

The system under consideration consists of an $Er^{3+}$:YAG crystal with a three-level ladder-type energy-level configuration, as shown schematically in Figure 1. The ground and excited states are labeled by $|1\rangle$, $|2\rangle$ and $|3\rangle$ with transition frequencies $\omega_{21}$ and $\omega_{31}$. A weak probe beam with frequency $\omega_p$ and wave vector $\vec{k}_p$ couples the $|1\rangle \rightarrow |2\rangle$ transition, while an intense control beam with frequency $\omega_c$ and wave vector $\vec{k}_c$ couples $|2\rangle \rightarrow |3\rangle$. In this cyclic configuration, symmetry breaking enables the otherwise forbidden $|1\rangle \rightarrow |3\rangle$ transition via a second-order nonlinear sum-frequency process, giving rise to a generated signal field with frequency $\omega_s = \omega_p + \omega_c$. Phase matching is satisfied when $\vec{k}_s = \vec{k}_c + \vec{k}_p$.

Under the electric-dipole and rotating-wave approximations, and neglecting counter-rotating terms, the interaction Hamiltonian of the system is expressed as ($\hbar=1$):

$$H = (\Delta_p |2\rangle\langle 2| + \Delta_p |3\rangle\langle 3|) - \frac{1}{2}(\Omega_p |2\rangle\langle 1| + \Omega_s |3\rangle\langle 1| + \Omega_c |3\rangle\langle 2| + c.c.) \qquad (1)$$

Here, $\Delta_p = \omega_{21} - \omega_p$ denotes the probe detuning, and the Rabi frequencies $\Omega_p$, $\Omega_c$, and $\Omega_s$ correspond to the probe, control, and generated beams, respectively.

Making use of the Liouville equation, we obtain the related density matrix equations of the system as follows:

$$\dot{\rho}_{11} = \frac{i}{2}(\Omega_p^* \rho_{21} - \Omega_p \rho_{12} + \Omega_s^* \rho_{31} - \Omega_s \rho_{13}) - \frac{1}{2}(\Gamma_{12}\rho_{21} + \Gamma_{21}\rho_{12} + \Gamma_{13}\rho_{31} + \Gamma_{31}\rho_{13}),$$

$$\dot{\rho}_{22} = \frac{i}{2}(\Omega_p \rho_{12} - \Omega_p^* \rho_{21} + \Omega_c^* \rho_{32} - \Omega_c \rho_{23}) - \frac{1}{2}(\Gamma_{21}\rho_{12} + \Gamma_{12}\rho_{21} + \Gamma_{23}\rho_{32} + \Gamma_{32}\rho_{23}),$$

$$\dot{\rho}_{33} = \frac{i}{2}(\Omega_s \rho_{13} - \Omega_s^* \rho_{31} + \Omega_c \rho_{23} - \Omega_c^* \rho_{32}) - \frac{1}{2}(\Gamma_{31}\rho_{13} + \Gamma_{13}\rho_{31} + \Gamma_{23}\rho_{32} + \Gamma_{32}\rho_{23}),$$

$$\dot{\rho}_{21} = -i(\Delta_p - i\gamma_{21})\rho_{21} + \frac{i}{2}\Omega_p(\rho_{11} - \rho_{22}) + \frac{i}{2}\Omega_c^* \rho_{31} - \frac{i}{2}\Omega_s \rho_{23}, \qquad (2)$$

$$\dot{\rho}_{31} = -i(\Delta_p - i\gamma_{31})\rho_{31} + \frac{i}{2}\Omega_s(\rho_{11} - \rho_{33}) + \frac{i}{2}\Omega_c \rho_{21} - \frac{i}{2}\Omega_p \rho_{32},$$

$$\dot{\rho}_{32} = -\gamma_{32}\rho_{32} + \frac{i}{2}\Omega_c(\rho_{22} - \rho_{33}) + \frac{i}{2}\Omega_s \rho_{12} - \frac{i}{2}\Omega_p^* \rho_{31},$$

$$\rho_{11} + \rho_{22} + \rho_{33} = 1$$

where $\gamma_{21} = (\Gamma_{21} + \gamma_2)/2$, $\gamma_{31} = (\Gamma_{31} + \Gamma_{32} + \gamma_3)/2$, and $\gamma_{32} = (\Gamma_{32} + \gamma_3)/2$. Here, $\Gamma_{ij}$ ($i, j = 1, 2, 3; i > j$) denotes the spontaneous decay rate from level $|i\rangle$ to level $|j\rangle$, while $\gamma_i$ stands for the non-radiative decay rate of level $|i\rangle$. Due to the multi-phonon decay of the crystal, $\Gamma_{31} \ll \gamma_3$ is obtained. Therefore, we can consider $\Gamma_{31} \approx 0$, $\Gamma_{22} + \gamma_2 \approx 1/\tau_2$ and $\Gamma_{33} + \gamma_3 \approx 1/\tau_3$ as an approximation. $\tau_2$ and $\tau_3$ are the lifetimes of the levels $|2\rangle$ and $|3\rangle$.

We use the weak-field approximation and apply the perturbation $\rho_{ij} = \rho_{ij}^{(0)} + \rho_{ij}^{(1)} + ...$, where the first term ($\rho_{ij}^{(0)}$) is recognized as the zeroth-order term, and the subsequent terms reflect the higher-order coherence terms representing linear and nonlinear optical processes [5]. Moreover, the three-level quantum system is assumed to be initially populated in the ground state $|1\rangle$. Based on these assumptions, the zeroth-order solution yields $\rho_{11}^{(0)} \approx 1$ in which the other elements vanish ($\rho_{ij}^{(0)} \simeq 0$). In the steady state of the beams, the off-diagonal density-matrix elements are computed analytically. These bipartite elements comprise the corresponding first-order and second-order processes, which are formulated as:

$$\rho_{21}^{(1)} = \rho_{21}^{(L)} + \rho_{21}^{(NL)} = \frac{i\Omega_p(i\Delta_p + \gamma_{31})}{2\xi} + \frac{i^2\Omega_c^*\Omega_s}{4\xi}, \tag{3}$$

$$\rho_{31}^{(1)} = \rho_{31}^{(L)} + \rho_{31}^{(NL)} = \frac{i\Omega_s(i\Delta_p + \gamma_{21})}{2\xi} + \frac{i^2\Omega_c\Omega_p}{4\xi}, \tag{4}$$

with $\xi = (i\Delta_p + \gamma_{21})(i\Delta_p + \gamma_{31}) + |\Omega_c|^2/4$. The first terms in both above equations ($\rho_{21}^{(L)}$ and $\rho_{31}^{(L)}$) describe linear absorption, while the second-order terms ($\rho_{21}^{(NL)}$ and $\rho_{31}^{(NL)}$) account for nonlinear three-wave mixing and reabsorption processes. For the sake of simplicity, the off-diagonal matrix elements can be recast as follows;

$$\rho_{21}^{(1)} = a_1\Omega_p + b_1\Omega_s, \tag{5}$$

$$\rho_{31}^{(1)} = a_2\Omega_p + b_2\Omega_s, \tag{6}$$

where $a_1 = i(\gamma_{31} + i\Delta_p)/2\xi$, $b_1 = i^2\Omega_c^*/4\xi$, $a_2 = i^2\Omega_c/4\xi$ and $b_2 = i(\gamma_{21} + i\Delta_p)/2\xi$.

Substituting Eqs. (5)–(6) into the Maxwell equations under the slowly varying envelope approximation yields the following propagation equations for the probe and signal beams:

$$\frac{\partial \Omega_p}{\partial z} = i\beta_{21}(a_1\Omega_p + b_1\Omega_s), \tag{7}$$

$$\frac{\partial \Omega_s}{\partial z} = i\beta_{31}(a_2\Omega_p + b_2\Omega_s), \tag{8}$$

Where $\beta_{21} = 2\pi N\omega_p|\mu_{21}|^2/C\hbar$ and $\beta_{31} = 2\pi N\omega_s|\mu_{31}|^2/C\hbar$ are the propagation constants. Here $\mu_{21}$ ($\mu_{31}$) denotes the dipole moment for the transition $|1\rangle \leftrightarrow |2\rangle$ ($|1\rangle \leftrightarrow |3\rangle$) and $N$ is the density of the electron [29]. Let us consider a scenario where the probe field is present at the entrance to the crystal $\Omega_p(Z=0) = \Omega_p(0)$ while the signal beam is absent $\Omega_s(Z=0) = 0$. After solving the coupled wave equations, Rabi frequencies $\Omega_s(Z)$ and $\Omega_p(Z)$ achieve:

$$\Omega_p(Z) = \left(\frac{A_1 - B_2}{A}\right)\Omega_p(0)\left[\sinh[AZ/2] + \frac{A}{A_1 - B_2}\cosh[AZ/2]\right]e^{\frac{A_1+B_2}{2}Z}, \tag{9}$$

$$\Omega_s(Z) = \frac{A_2}{A}\Omega_p(0)\sinh[AZ/2]e^{\frac{A_1+B_2}{2}Z} \tag{10}$$

where $Z = \beta_{21}z$ is the effective propagation distance. With constant coefficients defined as;

$$\begin{aligned} A_1 &= i\beta_{21}a_1, \quad B_1 = i\beta_{21}b_1 \\ A_2 &= i\beta_{31}a_2, \quad B_2 = i\beta_{31}b_2 \\ A &= \sqrt{(A_1 - B_2)^2 + 4A_2B_1} \end{aligned} \tag{11}$$

Eq. (9) represents the Rabi frequency of the probe field, and Eq. (10) characterizes the propagation of the generated signal beam in the medium. Note that the coherence term $\rho_{21}^{(1)}$ can be easily obtained by substituting Eqs. (9) and (10) into Eq. (5) and for $\rho_{31}^{(1)}$ by replacing them in Eq. (6):

$$\begin{aligned} \rho_{21}^{(1)} &= -\frac{a_1\Omega_p(0)}{2\xi A}\left((\gamma_{31} + i\Delta_p)\beta_{21} + (\gamma_{21} + i\Delta_p)\beta_{31}\right)\sinh[AZ/2]e^{\frac{A_1+B_2}{2}Z} \\ &\quad -\frac{A_2}{A}\Omega_p(0)\sinh[AZ/2]e^{\frac{A_1+B_2}{2}Z} + a_1\Omega_p(0)\cosh[AZ/2])e^{\frac{A_1+B_2}{2}Z}, \end{aligned} \tag{12}$$

$$\rho_{31}^{(1)} = a_2\Omega_p(0)\frac{A_1}{A}\left(\sinh[AZ/2] + \frac{A}{A_1}\cosh[AZ/2]\right)e^{\frac{A_1+B_2}{2}Z}. \tag{13}$$

It is worth mentioning that the analytical solutions for the Rabi frequencies of the probe and generated beams, as detailed in Eqs. (9) and (10) depend explicitly on the effective propagation distance and the $Er^{3+}$ concentration through the dipole moment and Rabi frequency parameters listed in Table 1. Hence, the concentration of $Er^{3+}$ ions strongly influence the optical response by modifying the dipole moment, level lifetimes, and Rabi frequencies of the control field, as summarized in Table 1.

In the rest of the paper, we assume the probe beam carries an optical vortex and OAM. Therefore, the Rabi frequency of the probe beam when it enters $Er^{3+}$-doped YAG crystals is written as:

$$\Omega_p(0) = E_p\left(\frac{r}{\omega}\right)^{|\ell|}e^{-\frac{r^2}{\omega^2}}e^{i\ell\phi}, \tag{14}$$

with $E_p$, r, $\omega$, $\ell$, and $\phi$ parameters denoting the strength of the beam, a cylindrical radius, the beam waist, the Orbital Angular Momentum (OAM) number, and the azimuthal angle [5]. Assuming that the direction of propagation of the ray is in the $Z$ direction, $r$ and $\phi$ in Eq. (13) equal with $\sqrt{x^2 + y^2}$ and $\phi = \tan^{-1}(y/x)$ respectively.

**Table 1.** Electric dipole moment, population decay rate, and Rabi frequency of control field in $Er^{3+}$: YAG crystal with different concentrations of the $Er^{3+}$ ion.

| concentration of the $Er^{3+}$ ion | Electric dipole moment $(10^{-32} Cm)$ | Level lifetime ($\mu s$) | | Rabi frequency of the control field $(10^{-20} Cm^2)$ |
|---|---|---|---|---|
| at % | $\mu_{23}$ | $^4I_{13/2}$ ($\tau_2$) | $^4S_{3/2}$ ($\tau_3$) | $\Omega_c$ |
| 0.5 | 1.69 | 1.00 | 2.15 | 25.35 |
| 3 | 1.84 | 1.38 | 1.61 | 27.60 |
| 15 | 1.36 | 1.38 | 0.15 | 20.40 |
| 33 | 1.19 | 0.31 | 0.08 | 17.85 |
| 100 | 1.14 | 0.04 | 0.01 | 17.10 |

## 3. Results and Discussion

To explore the optical vortex exchange and the role of $Er^{3+}$ concentration on coherence and propagation dynamics, we numerically analyze the coupled propagation equations (9)–(10) and the corresponding density matrix elements under steady-state conditions. The parameters of the $Er^{3+}$: YAG system used in all calculations are listed in Table 1.

### 3.1 Conversion Efficiency and Optimal Ion Concentration

To measure the efficiency of the three-wave mixing process in the system, we employ the conversion efficiency of the generated signal field, which is defined as [32]:

$$\eta = \left| \frac{E_s(Z)}{E_p(0)} \right|^2 = \left| \frac{\mu_{21}\Omega_s(Z)}{\mu_{31}\Omega_p(0)} \right|^2, \quad (15)$$

here $E_s(Z)$ is the amplitude (Rabi frequency) of the generated beam at arbitrary $Z$ and $E_p(0)$ ($\Omega_p(0)$) is the amplitude (Rabi frequency) of the probe field at the entrance $Z = 0$. By substituting Eq.s (9) and (10) in Eq. (14), $\eta$ obtains as follows:

$$\eta = \left( \frac{\beta_{31}^2 |\Omega_c|^2 \mu^2 \sinh^2\left(\frac{AZ}{2}\right)}{4\left(\beta_{21}(\gamma_{31} + i\Delta_p) + \beta_{31}(\gamma_{21} + i\Delta_p)\right)^2 - 12\beta_{31}^2 |\Omega_c|^2} \right) \exp\left[(A_1 + B_2)Z + (A_1^* + B_2^*)Z\right], \quad (16)$$

where $\mu = (\mu_{21}/\mu_{31})^2$. Moreover, by assuming $\gamma_{21}/\gamma_{31} = \beta_{21}/\beta_{31} = 3$ [29] one can obtain the simplified form of the conversion efficiency at $\Delta_p = 0$ as follows:

$$\eta = 3\mu^2 \sinh^2\left(\frac{i\beta_{21}\Omega_c Z}{8\gamma_{21}^2 + 6|\Omega_c|^2}\right) e^{-2\beta_{21}\gamma_{21}Z/3}. \quad (17)$$

Figure 2 presents the conversion efficiency ($\eta$) as a function of the effective propagation distance $Z$ for different $Er^{3+}$-ion concentrations. Each curve exhibits a characteristic rise followed by a decay, corresponding to the competition between nonlinear gain and absorption in the medium. For all concentrations, the efficiency initially increases with propagation distance as the nonlinear three-wave mixing process becomes dominant, then decreases due to reabsorption and phase-mismatch effects at larger $Z$. The maximum efficiency occurs at approximately $Z = 8.5$ for a concentration of 3%, as indicated by the dotted back line. This concentration offers the optimal balance between dipole strength and population coherence. At lower concentrations (0.5%), the coupling is too weak for efficient energy transfer, while higher concentrations (>15%) introduce dephasing and energy dissipation that suppress the nonlinear response. These results demonstrate that ion concentration acts as a tunable parameter controlling the effective nonlinear gain and the spatial evolution of vortex-beam transfer within the crystal.

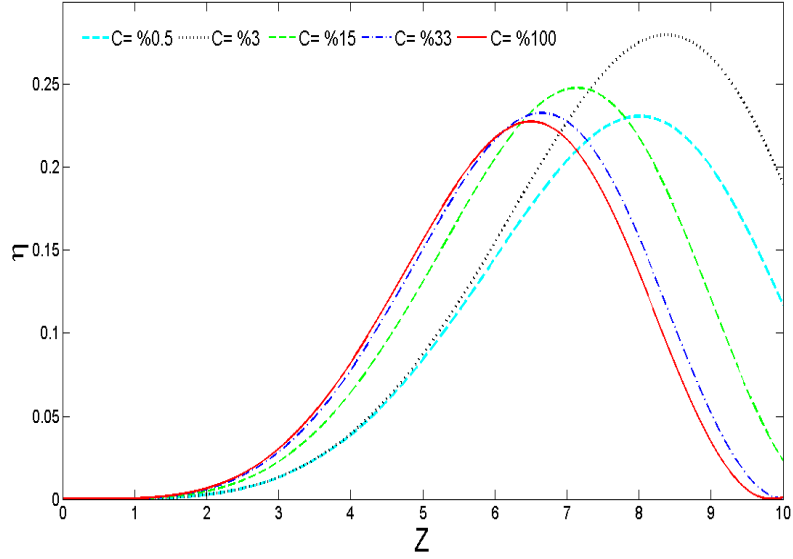

**Figure 2.** The conversion efficiency ($\eta$) versus effective propagation distance Z for different values of concentration of the $Er^{3+}$ ion. C=0.5% ($\Omega_c = 25.35$, $\mu_{21} = 3.25$, $\mu_{31} = 1.59$), C=3% ($\Omega_c = 27.60$, $\mu_{21} = 3.81$, $\mu_{31} = 1.72$), C=15% ($\Omega_c = 20.40$, $\mu_{21} = 2.82$, $\mu_{31} = 1.27$), C=33% ($\Omega_c = 17.85$, $\mu_{21} = 2.47$, $\mu_{31} = 1.11$), and C=100% ($\Omega_c = 17.10$, Values of the other parameters are taken as follows: $\Delta_p = 0$, $\Omega_p(0) = 5$, $\beta_{21} = 3\beta_{31} = 8$, $\ell = 1$ and $\gamma_{21} = 3\gamma_{31} = 3$.

Figure 3 shows the conversion efficiency as a function of $\Delta_p$ for various $Er^{3+}$ concentrations. For each concentration, the efficiency reaches a peak near resonance ($\Delta_p \approx 0$). Among all concentrations, the 3% $Er^{3+}$-doped sample again provides the highest peak efficiency, indicating minimal absorption. This behavior underscores the key interplay between detuning, coherence, and ion concentration in determining the overall nonlinear coupling strength. Small detuning enhances constructive interference between the probe and control fields, promoting efficient frequency conversion and OAM transfer. This figure highlights the significance of both the concentration of ions and the detuning of the probe beam in achieving the desired efficiency.

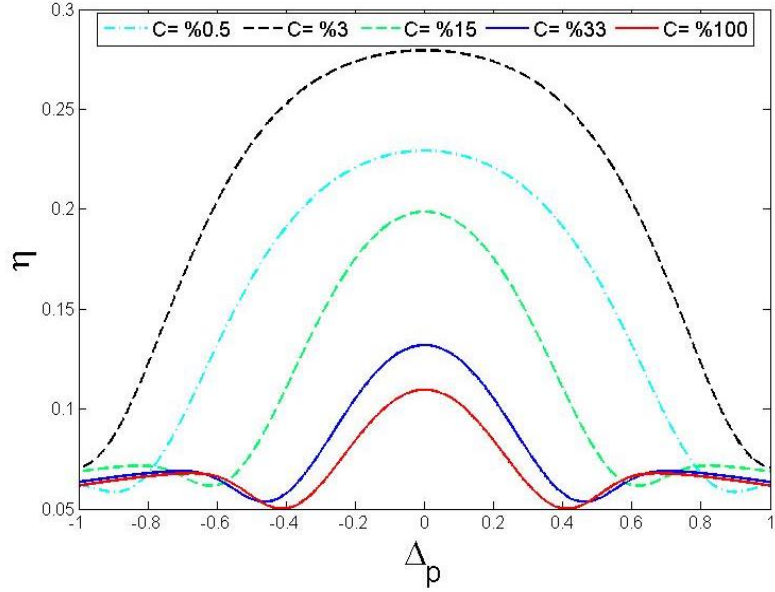

**Figure 3.** The conversion efficiency ($\eta$) versus the detuning of the probe field for $Z = 8.5$. Other parameters are the same as those in Fig. 2

## 3.3 Transfer of the Optical Vortex

The central goal of this work is to demonstrate the transfer of orbital angular momentum (OAM) from the probe beam to the generated signal beam. In the present model, the incident probe field is assumed to carry an optical vortex characterized by topological charge $\ell$, as described by Eq. (14). The generated signal field inherits this spatial phase structure through the nonlinear sum-frequency process governed by Eq. (10).

Figure 4 depicts the intensity and phase profiles of the generated signal beam as functions of the normalized positions $x/\omega$ and $y/\omega$ for OAM values $\ell = 0, 1, 2$ at the optimal concentration C = 3% and propagation distance Z = 8.5. When $\ell = 0$, the generated beam exhibits a Gaussian profile with a bright central maximum, corresponding to a non-vortex mode. For $\ell = 1$ and $\ell = 2$, the intensity distributions develop characteristic doughnut-shaped rings with dark cores, while the phase profiles display helical twisting around the beam axis. The number of phase windings increases linearly with $\ell$, confirming that the OAM of the probe beam is fully transferred to the generated signal beam. These results highlight the phase-preserving nature of the three-wave mixing process in this configuration. The strong coherence between the probe and control fields ensures that the phase information encoded in the vortex structure is conserved during energy conversion. This property is crucial for potential applications in quantum information transfer using OAM states.

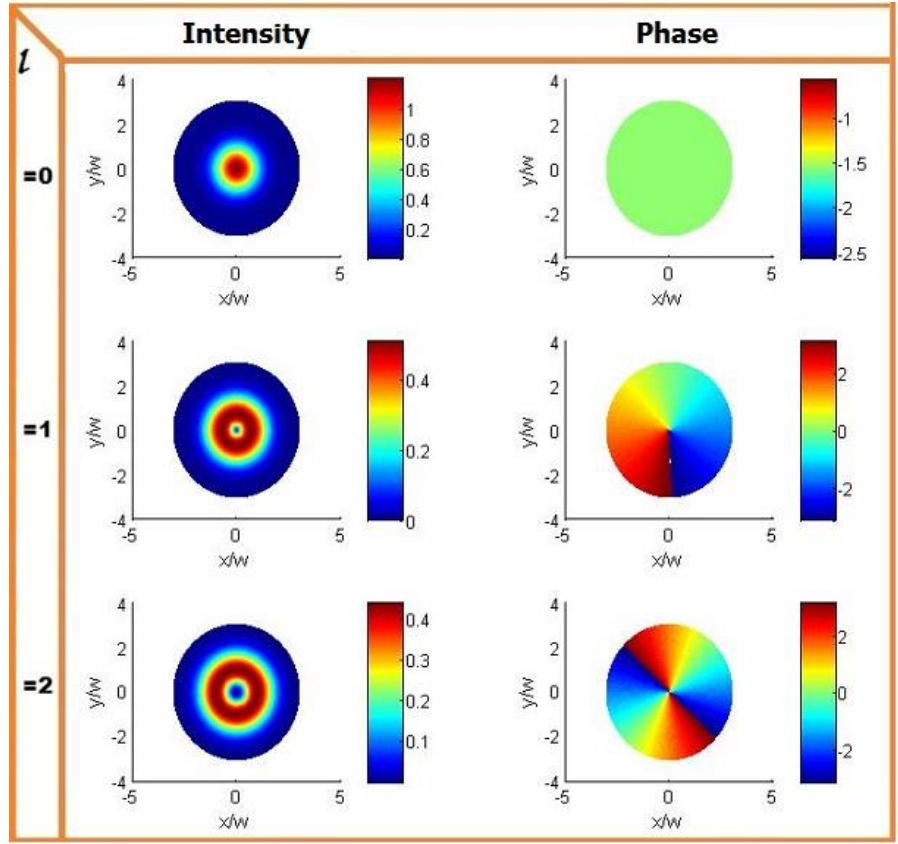

**Figure 4.** The intensity (left column) and phase (right column) of the generated signal beam as a function of the normalized positions $x/\omega$ and $y/\omega$. The analysis is for cases in which the probe possesses an optical vortex with orbital angular momentum values of 0, 1, and 2. The other parameters are set to $\gamma_{21} = 3\gamma_{31} = 3$, $\beta_{21} = 3\beta_{31} = 8$, $\Delta_p = 0$, $Z = 8.5$, $\Omega_p(0) = 5$ and C=3% ($\Omega_c = 27.60$).

## 3.3 Spatial Absorption Distribution

The complex susceptibility $\chi_{p(s)} = \chi'_{p(s)} + i\chi''_{p(s)}$ governs the optical properties of the probe (generated signal) beam, where the imaginary part $\chi''_{p(s)}$ describes absorption and the real part $\chi'_{p(s)}$ determines dispersion and refractive index modulation. It is well known that $\chi_{p(s)}$ is directly proportional to $\rho^{(1)}_{21}$ ($\rho^{(1)}_{31}$) with two terms: absorption (imaginary part) and dispersion (real part). By assuming $(\gamma_{21}/\gamma_{31}) = (\beta_{21}/\beta_{31}) = 3$ in Eqs. (12) and (13), we can express linear absorption and dispersion of the probe and the generated signal beams as follows:

$$\chi_p'' \propto \text{Im}[\rho_{21}^{(1)}] = \left(\frac{\alpha\beta_{21}}{12\xi^2}\right)\Omega_p(0)e^{\frac{A_1+B_2}{2}Z}\left\{\frac{|\Omega_c|^2}{4}+2\Delta_p^2\right\}\sinh[AZ/2]$$

$$+\frac{\gamma_{21}}{6\xi}\Omega_p(0)e^{\frac{A_1+B_2}{2}Z}\cosh[AZ/2] \qquad (18)$$

$$\chi_p' \propto \text{Re}[\rho_{21}^{(1)}] = \Omega_p(0)e^{\frac{A_1+B_2}{2}Z}\left\{-\left(\frac{\Delta_p}{2\xi}\right)\cosh[AZ/2]-\left(\frac{\beta_{21}\gamma_{21}\Delta_p}{18\xi^2 A}\right)\sinh[AZ/2]\right\} \qquad (19)$$

$$\chi_s'' \propto \text{Im}[\rho_{31}^{(1)}] = \left(\frac{1}{8\xi^2}\right)\Omega_c\Omega_p(0)e^{\frac{A_1+B_2}{2}Z}\left\{\frac{1}{3}\beta_{21}\gamma_{21}\alpha\sinh[AZ/2]-2\xi\cosh[AZ/2]\right\} \qquad (20)$$

$$\chi_s' \propto \text{Re}[\rho_{31}^{(1)}] = -\left(\frac{\alpha'\Delta_p\beta_{21}}{8\xi^2}\right)\Omega_c\Omega_p(0)e^{\frac{A_1+B_2}{2}Z}\sinh[AZ/2] \qquad (21)$$

where $\alpha' = 1/\sqrt{(-\beta_{21}(\gamma_{31}+i\Delta_p)+\beta_{31}(\gamma_{21}+i\Delta_p))^2-\beta_{31}\beta_{21}\frac{|\Omega_c|^2}{4\xi^2}}$ and

$\alpha = 1/\sqrt{-(4\Delta_p^2\beta_{21}^2/9)-(\beta_{21}^2|\Omega_c|^2/12\xi^2)}$ for the above equations.

To further understand the propagation dynamics, we plot the absorption profiles of both the generated signal and probe beams as functions of the normalized transverse coordinates $x/\omega$ and $y/\omega$ for different vortex modes in Figure 5. When the probe is a Gaussian beam ($\ell=0$), absorption is concentrated at the beam center, where the field intensity is highest. However, for nonzero OAM values ($\ell=1,2$), the absorption maps exhibit distinct ring-like patterns with alternating regions of gain and loss, reflecting the spatial redistribution of energy due to the vortex phase structure. The symmetry observed in these absorption maps indicates spatial energy exchange between the probe and generated beams. Locations of high absorption in the probe field coincide with amplification in the signal field, demonstrating the localized transfer of optical energy mediated by vortex-induced coherence. As $\ell$ increases, the number of absorption–gain spots grow, and their spatial separation widens, consistent with the expanding ring radius of higher-order vortex modes. These spatially dependent features are manifestations of the medium's structured coherence, in which the vortex beam's azimuthal phase influences the local polarization response and absorption characteristics.

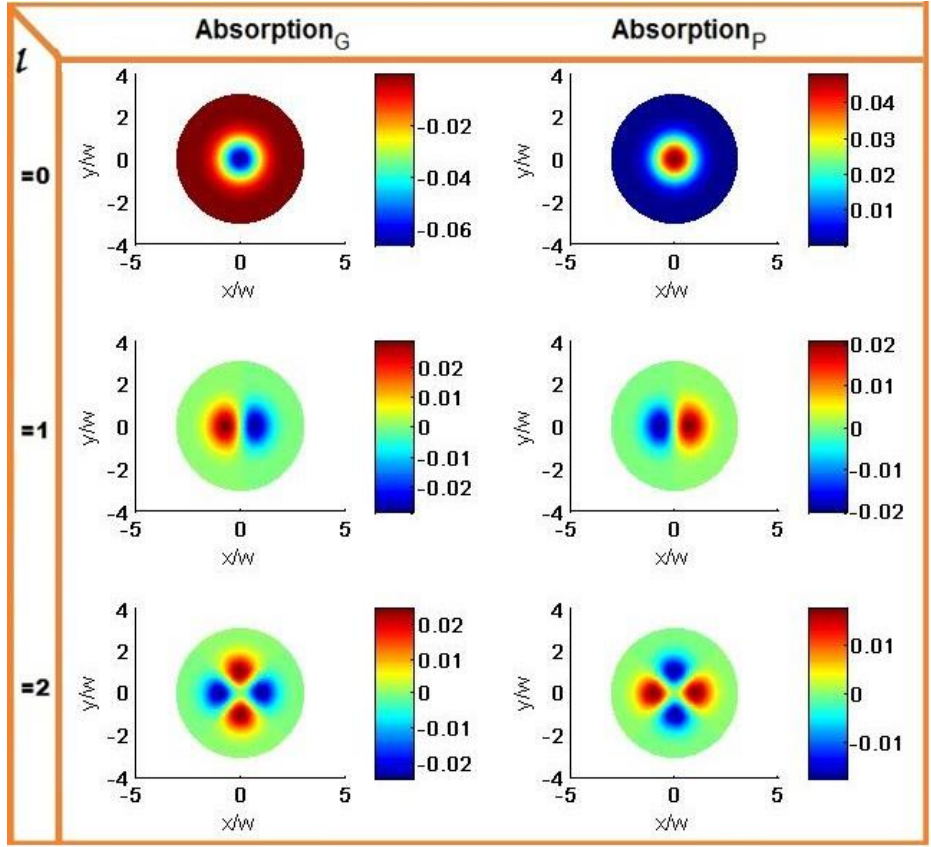

**Figure 5.** Absorption of the generated signal beam ($\mathrm{Im}[\rho_{31}^{(1)}]$) in the left column and absorption of the probe beam ($\mathrm{Im}[\rho_{21}^{(1)}]$) in the right column as functions of the normalized positions $x/\omega$ and $y/\omega$ for different modes of the probe LG beam $\ell = 0, 1, 2$. Other parameters are the same as those in Fig. 4.

### 3.4 Concentration-Dependent Absorption and Dispersion

We finally examine the effect of different concentrations of doped $Er^{3+}$ ions on the optical properties of the probe and generated beams.

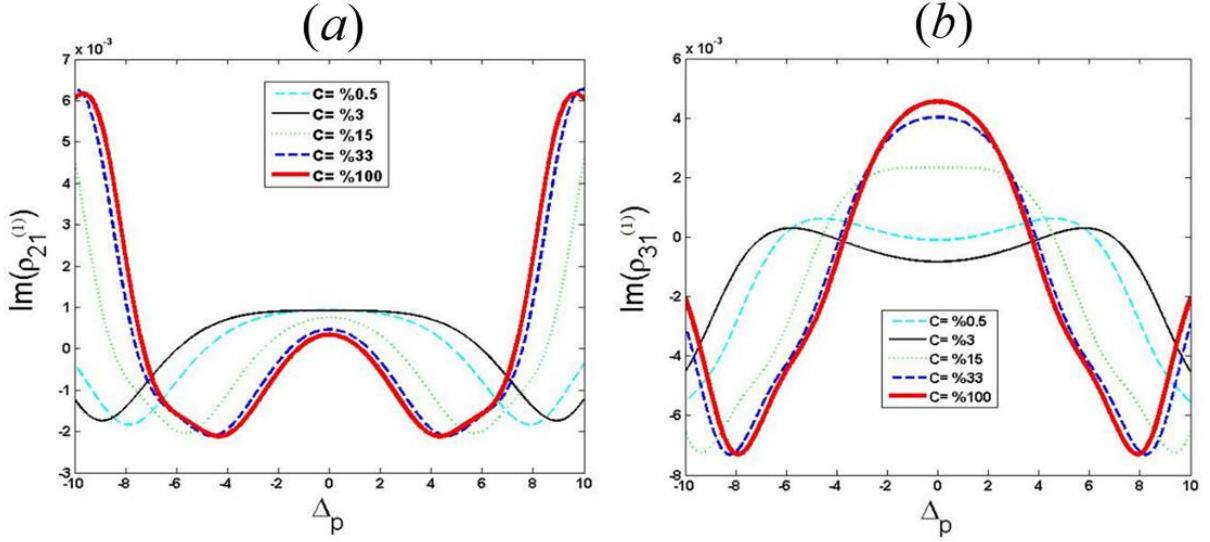

**Figure 6.** Plots of (a) $\text{Im}[\rho_{21}^{(1)}]$ (absorption of the probe beam) and (b) $\text{Im}[\rho_{31}^{(1)}]$ (absorption of the signal beam) as functions of detuning of the probe field ($\Delta_p$) for different values of the $Er^{3+}$ ion concentration. C=0.5% ($\Omega_c = 25.35$), C=3% ($\Omega_c = 27.60$), C=15% ($\Omega_c = 20.40$), C=33% ($\Omega_c = 17.85$), and C=100% ($\Omega_c = 17.10$). Values of the other parameters are taken as follows: $Z = 8.5$, $\Omega_p(0) = 0.1$ and $\ell = 1$.

Figure 6 shows the absorption spectra of (a) the probe and (b) the signal beams as functions of probe detuning $\Delta_p$. The probe absorption (Fig. 6a) shows a pronounced concentration dependence. Specifically for $-4 < \Delta_p < 4$, at low concentration (0.5%), the coupling is weak and a prominent absorption peak appears. Increasing the concentration to 3% enhances coherent coupling and broadens the absorption peak. Further increasing the concentration (15% → 33%) narrows the absorption spectral peak and decreases on-resonance absorption relative to the 3% case. Consequently, at resonance, an EIT-like transparency window emerges, while in the off-resonant cases, gain occurs. Concentrations above ≈33% produce only minor additional changes, indicating that the optical response is approaching saturation. The absorption spectra of the generated signal beam (Fig. 6b) in the range are as follows: At a low concentration (0.5 %), the absorption around the resonance is almost zero, indicating limited transparency and weak coherence, while it increases slowly as one moves away from the resonance. As the concentration increases to 3%, the absorption crosses zero and becomes negative, indicating net amplification where energy is coherently transferred from the pump fields to the signal. Beyond this optimum, at higher concentrations (15%–33%), the spectra show net absorption in the resonance and near the resonance. Beyond 33 % concentration, the spectral shape stabilizes, indicating that the system has entered a saturation regime in which further doping does not appreciably alter the optical response.

Thus, for resonance and near-resonance conditions, the 3% concentration offers the best compromise for the probe, providing maximum absorption while the signal benefits, with efficient generation and possible localized gain.

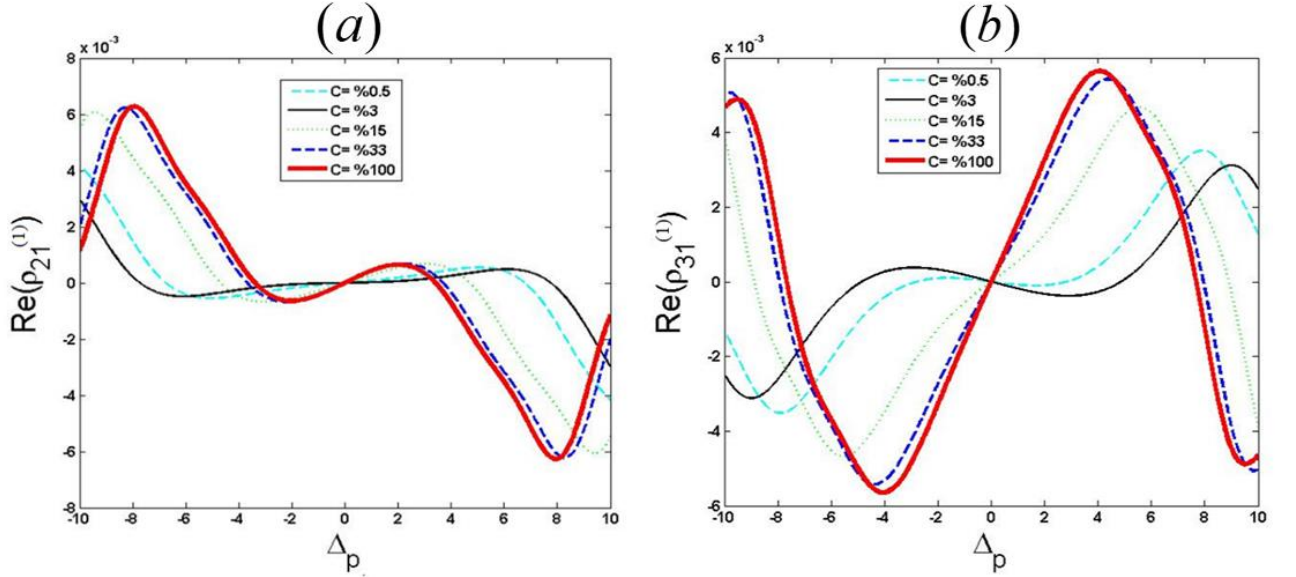

**Figure 7**. Plots of (a) $\text{Re}[\rho_{21}^{(1)}]$ (dispersion of the probe beam) and (b) $\text{Re}[\rho_{31}^{(1)}]$ (dispersion of the signal beam) as functions of detuning of the probe beam ($\Delta_p$) under various concentrations of $Er^{3+}$ ions in the YAG crystal.. Other parameters are the same as those in Fig. 6.

Figure 7 presents the dispersion spectra, that is, the real parts of the susceptibilities $\chi'_p$ and $\chi'_s$ for the (a) probe and (b) generated signal beams as functions of the probe beam ($\Delta_p$) for different values of the $Er^{3+}$ ion concentration. It is well-known that the group velocity of light propagation in optical media is governed by $v_g = c(n_r + \omega dn_r/d\omega)^{-1}$ in which the real part of the refractive index is defined as $n_r \propto 1 + \chi'$. Hence, a positive slope $d\chi'/d\omega$ corresponds to normal dispersion and subluminal (slow) light, while a negative slope for $d\chi'/d\omega$ corresponds to anomalous dispersion and potentially superluminal (fast) propagation. Here, too, we are interested in analyzing behavior in the range $-4 < \Delta_p < 4$. For the probe (Fig. 7(a)), the dispersion at low concentration (0.5%) changes slowly, with a slight positive slope relative to delta, resulting in a weak frequency dependence of the refractive index. When the concentration increases to 3%, the positive slope $d\chi'/d\Delta_p$ becomes smaller, and the dispersion value approaches zero, resulting in a group velocity that is nearly equal to the speed of light in a vacuum. Further increasing the Er$^{3+}$ concentration to 15–33% gradually steepens the dispersion curve, indicating that the refractive index changes more sharply with $\Delta_p$. As a result, the group velocity decreases, and the slow-light effect becomes stronger. Beyond 33%, the dispersion slope saturates, indicating that the optical response has reached a steady state in which additional doping no longer alters the refractive-index gradient. The signal-beam dispersion in Fig. 7(b) follows a different trend. We observe that increasing the concentration from 0.5% to 3% increases the negative dispersion slope, leading to higher group velocity and the possibility of superluminal propagation. Interestingly, a further increase in concentration reverses the slope of the dispersion curve, leading to a decrease in the speed of light and subluminal propagation.

**Conclusion**

In summary, we have theoretically investigated the coherent transfer of optical orbital angular momentum and dispersion-controlled light propagation in an $Er^{3+}$: YAG three-level ladder system. By solving the Maxwell–Bloch equations under the weak-probe and steady-state approximations, we obtained analytical expressions for the probe and generated beams that explicitly incorporate the effect of $Er^{3+}$-ion concentration. Our analysis reveals that the sum-frequency nonlinear process in this solid-state medium facilitates the efficient transfer of the optical vortex carried by the probe beam to the generated signal beam. The OAM transfer was examined for several topological charges, and in each case the generated field reproduced the characteristic doughnut-shaped intensity distribution and helical phase structure of the incident vortex beam, confirming complete transfer of the phase singularity and topological charge.

We further analyzed the dependence of vortex-conversion efficiency on $Er^{3+}$ concentration and showed that the transfer efficiency reaches a maximum at 3% doping, where coherent coupling is strongest. To elucidate the physical mechanism behind the vortex transfer, we examined the spatial absorption behavior and the frequency-dependent absorption spectra of both the probe and generated beams for different concentrations. The complementary absorption signatures—probe absorption suppression accompanied by signal amplification at optimal concentration—verify that energy and phase information are coherently redistributed between the two beams during OAM transfer. This behavior highlights the crucial role of concentration-dependent dispersion, coherence, and nonlinear coupling in facilitating vortex conversion on this solid-state platform. The tunable dispersion profile obtained from the real part of the susceptibility reveals a controllable transition between fast and slow propagation regimes. This feature demonstrates that $Er^{3+}$: YAG serves not only as a medium for structured-light conversion but also as an adjustable dispersive environment for coherent light-matter interactions.

These results position $Er^{3+}$: YAG crystals as a promising solid-state medium for the coherent manipulation of vortex beams and the propagation of structured light. The demonstrated vortex-transfer capability and dispersion control may support high-dimensional quantum communication and information processing using OAM-encoded photons [33,34], wavelength conversion of structured light for photonic interface compatibility [35], and slow-light-based coherent optical buffering in rare-earth-doped media [36,37]. The combination of OAM preservation and concentration-tunable dispersion suggests a path toward compact solid-state devices for structured-photon control and hybrid quantum interfaces.